\documentclass[%
 reprint,
superscriptaddress,
 amsmath,amssymb,
 aps,
floatfix,
]{revtex4-2}

\usepackage[caption=false]{subfig}
\usepackage{graphicx}
\usepackage{dcolumn}
\usepackage{bm}

\usepackage{amssymb}
\usepackage{amsthm}

\usepackage{lineno}
\usepackage{xcolor}
\usepackage{natbib}

\begin{document}

\preprint{APS/123-QED}

\title{First experience with He conditioning of an SRF photoinjector}

\author{I.~Petrushina}
 \email{ipetrushina@bnl.gov}
 
\affiliation{%
Department of Physics and Astronomy, Stony Brook University, Stony Brook, NY
}%

\author{Y.~Jing}
\affiliation{Collider-Accelerator Department, Brookhaven National Laboratory, Upton, NY}%

\author{V.N.~Litvinenko}
\affiliation{%
Department of Physics and Astronomy, Stony Brook University, Stony Brook, NY
}%
\affiliation{Collider-Accelerator Department, Brookhaven National Laboratory, Upton, NY}

\author{J.~Ma}
\author{F.~Severino}
\author{G.~Narayan}
\author{T.~Hayes}
\author{J. C.~Brutus}
\author{L.~Smart}
\author{K.~Decker}
\affiliation{Collider-Accelerator Department, Brookhaven National Laboratory, Upton, NY}%

\author{S.~Belomestnykh}
\affiliation{Fermi National Accelerator Laboratory, Batavia, IL}%

\date{\today}

\begin{abstract}

The recent achievements in the performance of superconducting RF (SRF) photoinjectors have opened a new era in the development of the reliable high-brightness CW electron sources. While the SRF guns become one of the most promising technologies, the compatibility of SRF environment with the complex photocathodes remains on the forefront of the modern accelerator science. The SRF cavities operate at cryogenic temperatures providing the ultra-high vacuum (UHV) environment highly beneficial for the photocathode performance. However, the necessity to keep the photocathodes at room temperature while being surrounded by the cavity walls that are kept at cryogenic temperatures creates an additional complexity for the SRF gun design and operation. The complex and volatile chemical compounds used for photocathodes have a high chance of contaminating the surfaces of an SRF cavity. When deposited, such compounds could create centers for cold electron emission which degrade performance of the SRF guns. Such a circumstance would require development of the \emph{in-situ} processing techniques for restoring the SRF cavity performance. This paper presents the results of the successful implementation and application of the He conditioning method for cavity restoration using the existing SRF photoinjector at Brookhaven National Laboratory (BNL). The method has proven to be extremely effective and resulted in a dramatic improvement of the BNL gun performance.
\end{abstract}

\maketitle


\section{Introduction}
\label{sec1:Introduction}

The future development of modern X-ray Free Electron Lasers (XFELs), $\gamma$-ray sources, fixed target, and collider experiments for Nuclear and High Energy Physics (NP\&HEP), and a variety of other scientific, industrial, and medical applications, rely heavily on the ability to produce low emittance and high current CW electron beams \cite{Wang2016}. The development of reliable electron sources has been a major enabling technology for achieving such beams. The recent progress in the field of SRF photoinjectors have demonstrated a successful performance confirming that, indeed, the SRF environment and high quantum efficiency (high-QE) photocathodes are compatible and can produce high-brightness beams with low transverse emittances \cite{Petrushina2020}. Nevertheless, the nature of the photoinjector operation requires systematic cathode exchange, which puts the cleanliness of the photoinjector cavity at risk. A reliable \emph{in-situ} cavity restoration procedures can be of great importance when it comes to the cavity performance and maintenance.

A successful performance of an SRF photoinjector is a delicate balance of maintaining the good quality of operation for both the photocathodes and the cavity itself. The introduction of a photocathode into an SRF environment comes with a great risk of the cavity contamination by the particulates produced during the process of the cathode insertion/removal, and deposition of volatile or highly emissive compounds from the cathode (Cs, Na, K, etc.) onto the walls of the cavity during operation. Such contaminants create emission centers on the cavity surface and result in a significant cavity performance degradation caused by excessive production of \emph{dark current} and enhanced strength of \emph{multipacting} barriers. 

\emph{Dark current} is an undesirable emission of electrons from the regions of high electric fields on the cavity surface caused by a particulate contamination. An excessive dark current results in an increased generation of X-rays, higher consumption of liquid He, and, possibly, quenching of the cavity. By aggravating the dark current, the presence of a photocathode within the cavity body increases the risk of the reduced operational accelerating voltage of the photoinjector, and subsequent degradation of the beam quality.

The resonant build-up of the secondary electrons within the cavity body---\emph{multipacting} (MP)---strongly depends on the Secondary Emission Yield (SEY) of the cavity wall material and is directly impacted by the presence of external contaminants. MP prevents the cavity from reaching operational voltage, causes local overheating of the cavity surface, and may result in the cavity quenching. 

Both phenomena described above demonstrate an example of the damage to the cavity coming from the introduction of the photocathode into an SRF environment. On the other hand, if a cavity has a presence of strong MP barriers, the occurring electron bombardment can significantly damage the cathode surface and dramatically decrease the cathode lifetime. To address the issue of the inevitable contamination of the cavity walls, the reliable in-situ cavity restoration techniques must be developed. A variety of surface treatment techniques applicable for SRF cavities have been developed throughout the years, but their applications to the restoration of the SRF photoinjectors are still under investigation. 

In this paper we present our experience with handling of the 113 MHz SRF photoinjector with a warm $\textrm{CsK}_{2}\textrm{Sb}$ cathode throughout several years of operation. We report the complications encountered throughout these years and the solutions that were found to overcome these challenges. First, we will provide a brief overview of a few major SRF cavity restoration techniques available to date in application to the SRF accelerating structures. We will then proceed by providing a description of the 113 MHz SRF gun cavity at Brookhaven National Laboratory (BNL) and its incredible performance in Section~\ref{sec:gun}. Section~\ref{sec:HeC} describes our experience in application of the He conditioning technique for restoration of the BNL SRF gun performance and reports on the progress achieved using this treatment.

\section{Cavity restoration techniques}
\label{sec:restorationTechniques}

With the variety of restoration techniques applicable for the SRF cavities, there are a few that currently stand out when it comes to the prevention of the field emission: High Power Processing (HPP), He processing and plasma treatment \cite{padamsee1998,berrutti2017plasma}. These methods are generally implemented as \emph{in-situ} restoration techniques and have demonstrated significant improvement in the cavity performance.

\textbf{CW RF processing or Cold Conditioning (CC)} technique is fairly straightforward and involves operation of the cavity in a CW mode for an extended period of time while monitoring the RF losses and X-ray radiation levels from the cavity. An RF conditioning in a sense is quite similar to a DC breakdown when an applied surface electric field causes an explosive event resulting in vaporization of a surface emitter. The event of an emitter explosion is accompanied by a rapid drop in the field emission current, RF losses, radiation activity, and results in an immediate improvement of the cavity performance that reaches higher operational voltage levels.

\textbf{High Power Processing (HPP)} is a natural second step in the cavity conditioning after the CC method has reached its limits. HPP involves pulsed operation of the cavity which is beneficial for the emitter vaporization due to the ability to achieve higher surface electric fields at a reduced power dissipation and liquid helium consumption compared to the CC. 

\textbf{Helium Conditioning (HeC)} is usually performed at the operational cryogenic temperatures of the cavity by filling the cavity volume with pure He gas at pressures ranging from 10$^{-6}$ to 10$^{-5}$ Torr. Subsequent application of short high voltage RF pulses creates electric field at the surface of the dark current emitters above the critical level. When such high electric fields are applied to the impurity, the emitted burst of electrons ionizes the surrounding He gas, and He ions bombard the source and efficiently destroy such emitters.

\textbf{Plasma Treatment (PT)} of the SRF cavities is a relatively new method, and is actively pursued by the Fermi National Accelerator Laboratory (FNAL) team in application to the 1.3 GHz LCLS-II SRF cavities. PT is performed at room temperature and involves filling the cavity with a mixture of noble gases (such as Ne or Ar) and oxygen at pressures of 70-200 mTorr followed up with an ignition and sustenance of an RF discharge. A low percentage of oxygen is added to react with the hydrocarbons adsorbed on the cavity surface. The volatile byproducts are pumped out of the cavity and monitored with a Residual Gas Analyzer (RGA). Ne and O$_2$ are mixed before reaching the cavity and their ratio is controlled using an RGA. The \emph{in-situ} PT technique was developed mostly to mitigate hydrocarbon related field emission \cite{padamsee1998}. Starting from the SNS experience \cite{Doleans2016Ig,Doleans2016In} and using the new ignition method, PT has been applied to multiple 1.3 GHz cavities at FNAL and extended to using HOM (Higher Order Modes) couplers to ignite the glow discharge with only a few watts of RF power required \cite{berrutti2017plasma,berrutti2018update}. In multi-cell cavities, the plasma can be ignited one cell at a time using a superposition of HOMs. 

When the appropriate restoration methods are applied, the performance of an SRF cavity can be significantly improved resulting in a higher achieved electric fields, lower consumption of liquid He, and lower radiation levels. The investigation of application of such methods to SRF photoinjectors is crucial for the advancement in the development of novel SRF electron sources.

\section{113~MHz SRF gun}
\label{sec:gun}

\begin{figure*}
    \centering
    \includegraphics[width=0.7\linewidth]{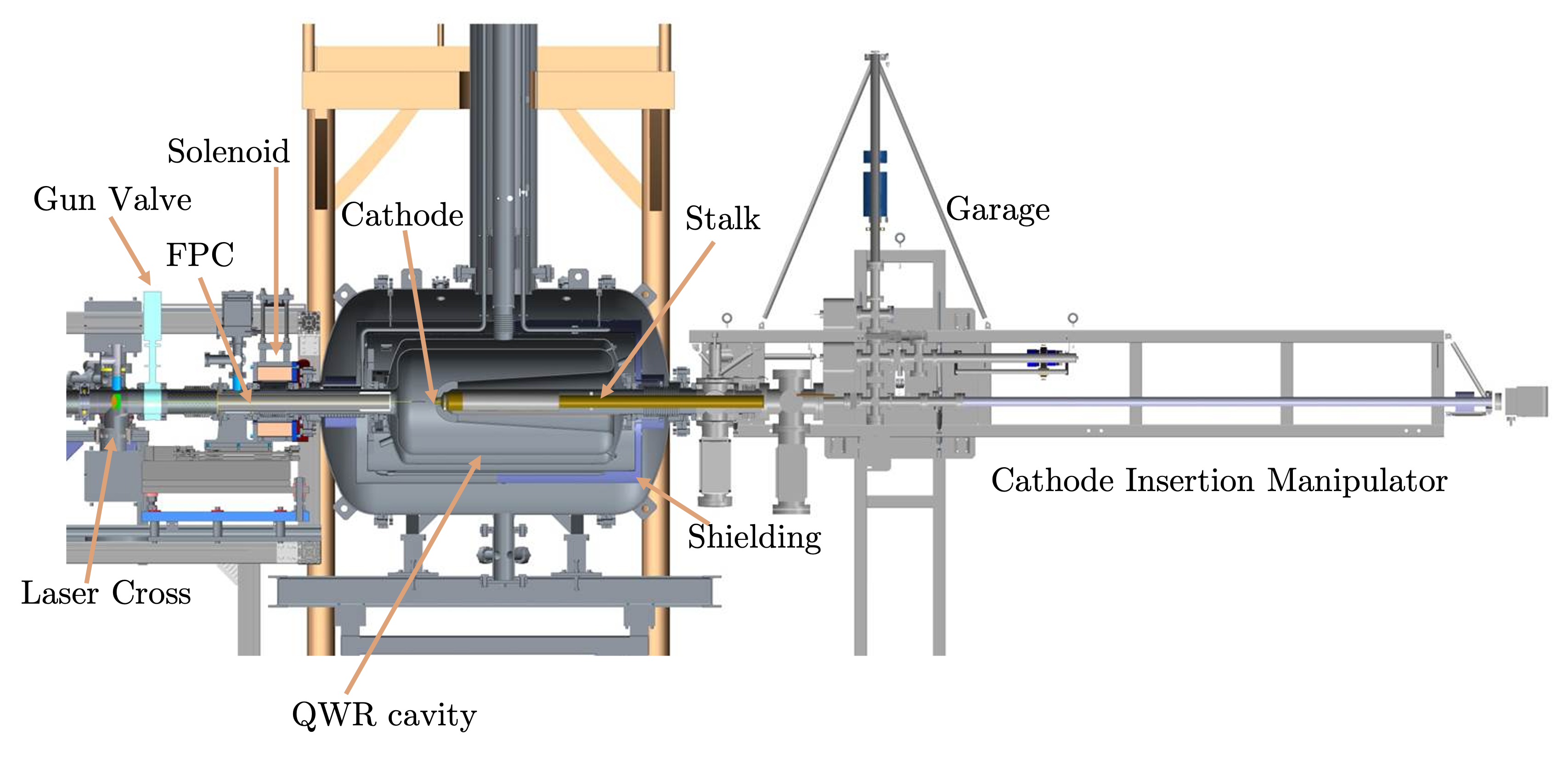}
    \caption{Schematic of the CeC SRF gun assembly indicating the major system components.}
    \label{fig:SRFgunAssembly}
\end{figure*}

\begin{figure*}
    \centering
    \includegraphics[width=1\linewidth]{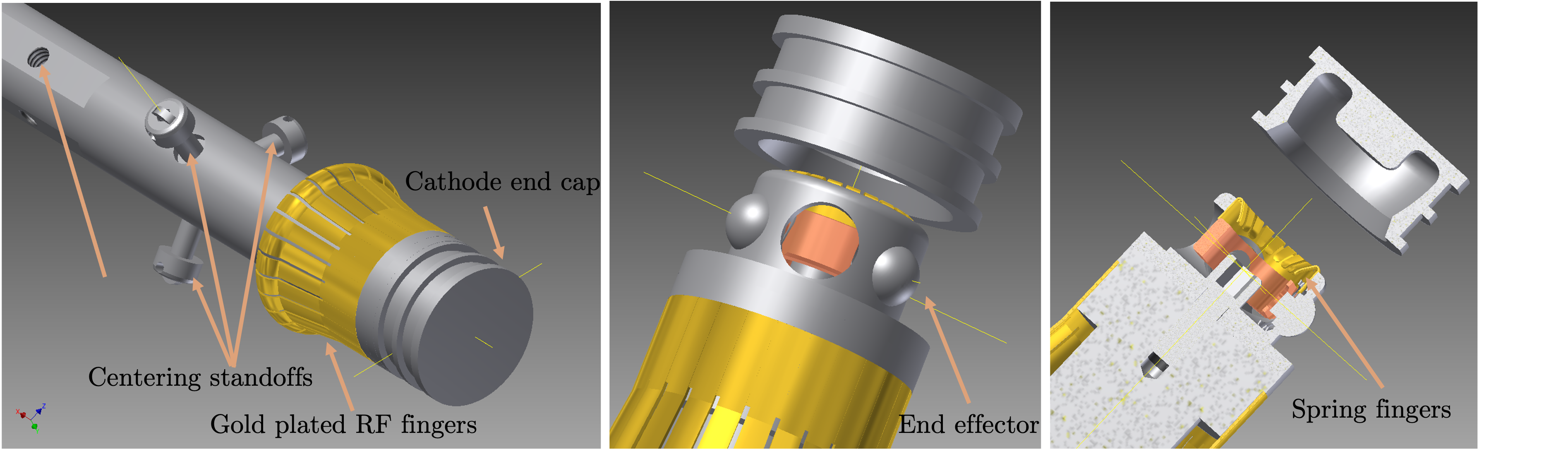}
    \caption{Detailed schematic of the cathode puck and cathode manipulator end effector assembly.}
    \label{fig:EndEffector}
\end{figure*}

The 113 MHz SRF gun was built for the Coherent electron Cooling Proof of Principle (CeC PoP) \cite{litvinenko2009coherent} experiment at BNL, and is a VHF-type quarter-wave resonator (QWR) made of bulk Niobium. The gun operates at a very low frequency which provides the beneficial condition for photoelectrons to be generated at the peak of the accelerating field (10-20~MV/m at the photocathode surface). While the detailed design of the cavity body and its auxiliaries can be found in \cite{pinayev2020high,belomestnykh2015commissioning,brutus2013mechanical}, here we will limit the description only to a general system overview with the focus on the specifics of the cathode delivery system as it's critical for the purpose of this paper. The full assembly of the QWR cavity in the cryostat with its surroundings is shown in Fig.~\ref{fig:SRFgunAssembly}. The cavity is maintained at 4~K and uses the liquid He (LiHe) from the refrigeration systems of the Relativistic Heavy Ion Collider (RHIC). The green 532~nm laser light is transported to the cathode surface through the laser cross and the hollow fundamental power coupler (FPC) that serves two purposes---delivers power from the 4~kW solid-state amplifier to the cavity and acts as a frequency tuner. The room temperature hollow Cu-coated stainless steel stalk resides inside the 4~K SRF gun. The stalk has an impedance transformer and functions as a half-wave choke that is shorted outside the cryostat. The cathode is delivered inside the cavity through the stalk with the use of the magnetic cathode manipulator system. The cathode delivery system has three UHV arms used to transport the cathode pucks from a storage unit called ``garage'' into the cavity. The long manipulator arm delivers the cathode to the end of the stalk, where it is grounded to the stalk by the RF spring fingers (the details of the cathode and cathode manipulator end effector are illustrated in Fig.~\ref{fig:EndEffector}). Controlling the depth of the stalk insertion with respect to the cavity nose allows for the control of the primary beam focusing near the cathode surface, which significantly improves the beam quality at the exit of the gun.

The $\textrm{CsK}_{2}\textrm{Sb}$ photocathode material is deposited onto the polished surface of Mo puck using a dedicated cathode deposition system. The garage, equipped with mobile UHV ion and sublimation pumps, can be attached to the system via a load-lock, and can hold up to three photocathodes at a time. When it is detached, the garage is transported to RHIC tunnel, where it is connected to the gun’s load-lock. After a brief bake-out, the cathodes can be transferred into the gun. Overall, the cathode exchange between the gun and the garage takes about 30 minutes.

The cavity was manufactured by Niowave, but has never achieved its design voltage of 2~MV due to the limitations during the production process. The routine CeC PoP operation requires the gun to deliver 1.25~MeV electron beams with a headroom of 200~kV, which was easily achievable. Overall, the cavity has shown an impeccable performance by sustaining CW operation during the CeC experimental runs (continuous use of the gun for 5-6 months each year). The early stages of operation were challenged by the presence of persistent low-level MP barriers, but the issue was methodically evaluated and successfully resolved resulting in complete elimination of MP as an obstacle for the cavity operation. The detailed description of our MP studies, and the dedicated start-up procedure developed to avoid the MP issue can be found in \cite{Petrushina2018}.

The cavity performance as a photoinjector has been rather impressive: the gun demonstrated the ability to deliver high-brightness electron bunches with low transverse emittances (e.g. 0.15~mm-mrad slice emittance for 100 pC, 400 ps bunches), and has also generated record high charges per bunch (up to 20~nC) \cite{Petrushina2020}. The cavity employs high quantum efficiency (QE) room temperature CsK$_{2}$Sb photocathodes, and has demonstrated the ability to maintain the cathode's QE of 3-4\% for several months of continuous operation \cite{Wang2021}.

\begin{figure}
    \centering
    \includegraphics[width=1\linewidth]{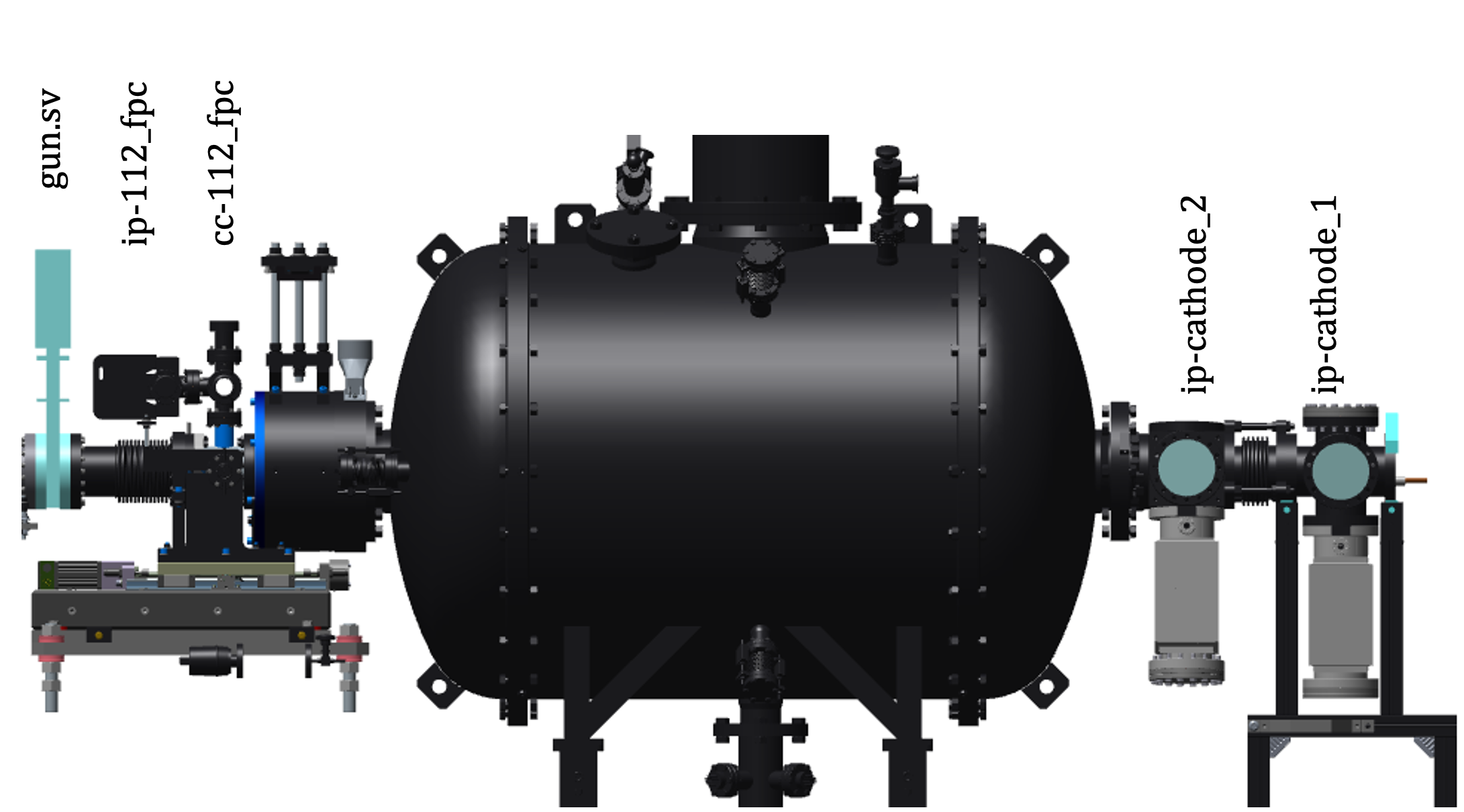}
    \caption{Detailed view of the SRF gun cryostat with the critical vacuum components. The cathode insertion is on the right side of the schematic, and the FPC is on the left side. Here ip = ion pump; tc = thermal conductivity gauge, reads pressure from atmosphere to 1e-3 Torr; cc = cold cathode gauge, reads 1e-3 to 1e-11 Torr; sv = vacuum valve. }
    \label{fig:GunDewarValvesAndPumps}
\end{figure}

\begin{figure}
    \centering
    \includegraphics[width=1\linewidth]{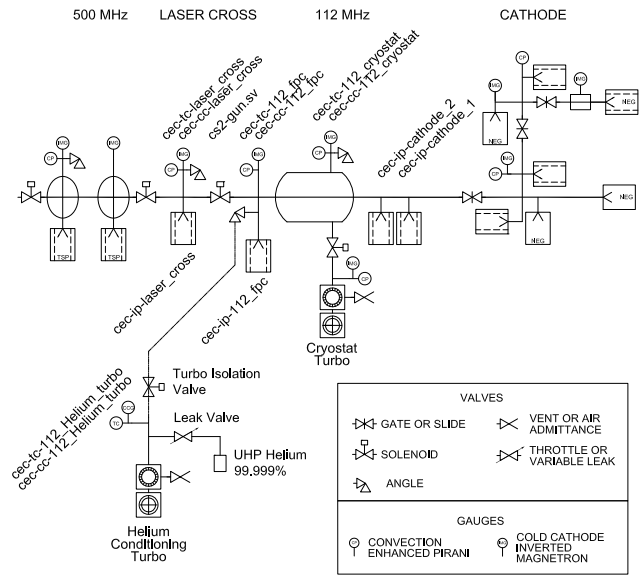}
    \caption{Schematic of the clean He gas delivery system for the conditioning of the CeC SRF cavities.  Here ip = ion pump; tc = thermal conductivity gauge; cc = cold cathode gauge; sv = vacuum valve; NEG = non-evaporable getter pump; UHP = ultra high purity grade.}
    \label{fig:CeCValvesAndPumps}
\end{figure}

\section{He Conditioning in application to the BNL SRF gun} 
\label{sec:HeC}

During the multiple years of the gun operation, we have learned that development of \emph{in-situ} methods of cavity performance restoration is crucial for the routine photoinjector maintenance. There are always multiple sources of possible SRF gun cavity contamination, including particulates generated by inserting and removing cathodes or those migrating from other parts of the vacuum system, which can result either in reduced performance of the SRF gun or even in loss of valuable piece of equipment. One of the successful methods that we have implemented for improving performance of our SRF gun is He conditioning (HeC). The system for HeC of the CeC SRF accelerator, designed and built by the BNL vacuum group, is shown in Figs.~\ref{fig:GunDewarValvesAndPumps} and \ref{fig:CeCValvesAndPumps}. Below we will discuss two cases when the HeC procedure was used in application to the BNL SRF gun and demonstrated a dramatic improvement in the cavity performance.

\subsection{He Conditioning: Case 1}

\begin{figure}
    \centering
    \subfloat{
    \includegraphics[height=4cm,width=.49\linewidth]{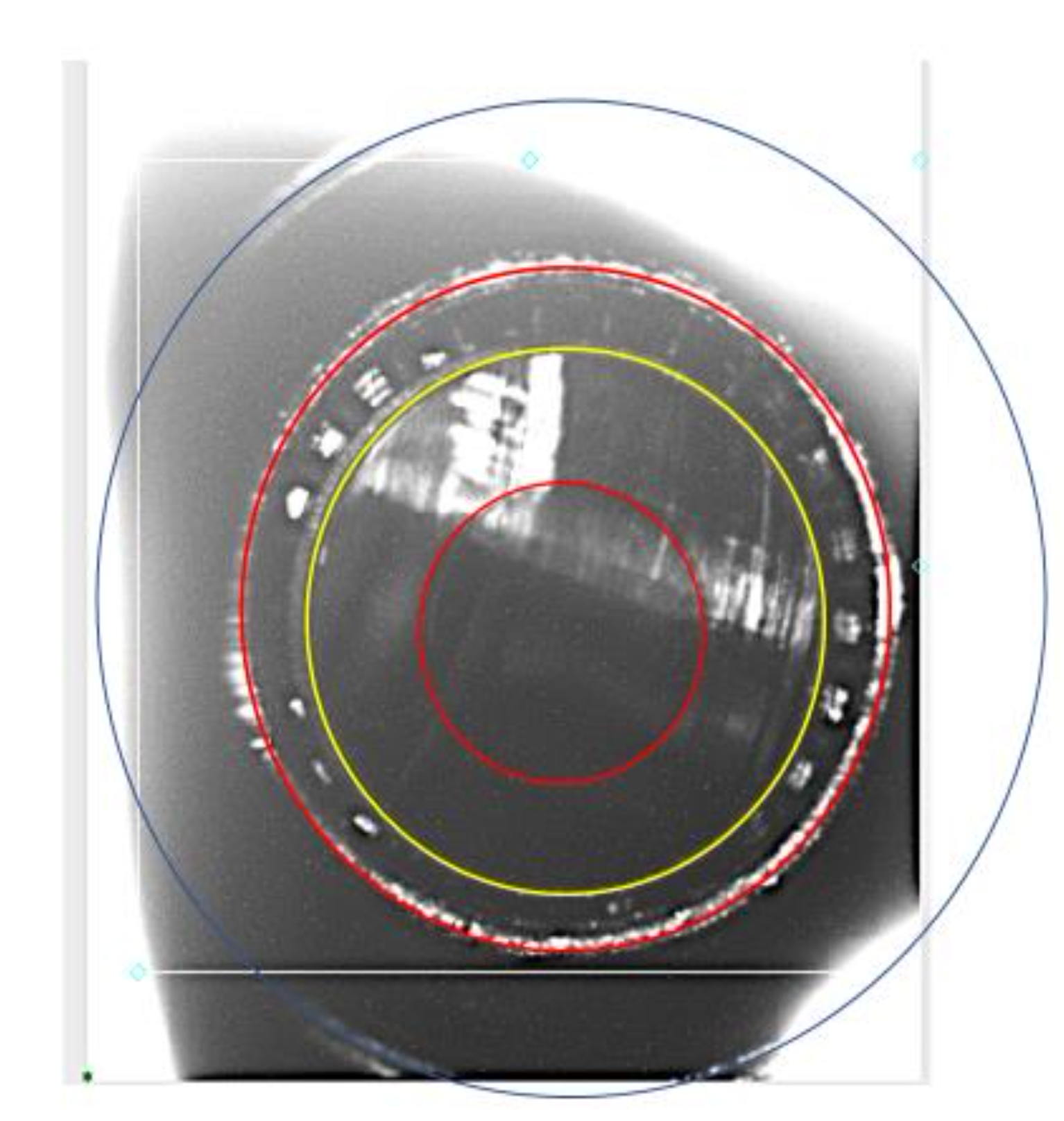}}
    \hfill
    \subfloat{%
    \includegraphics[height=3cm,width=.49\linewidth]{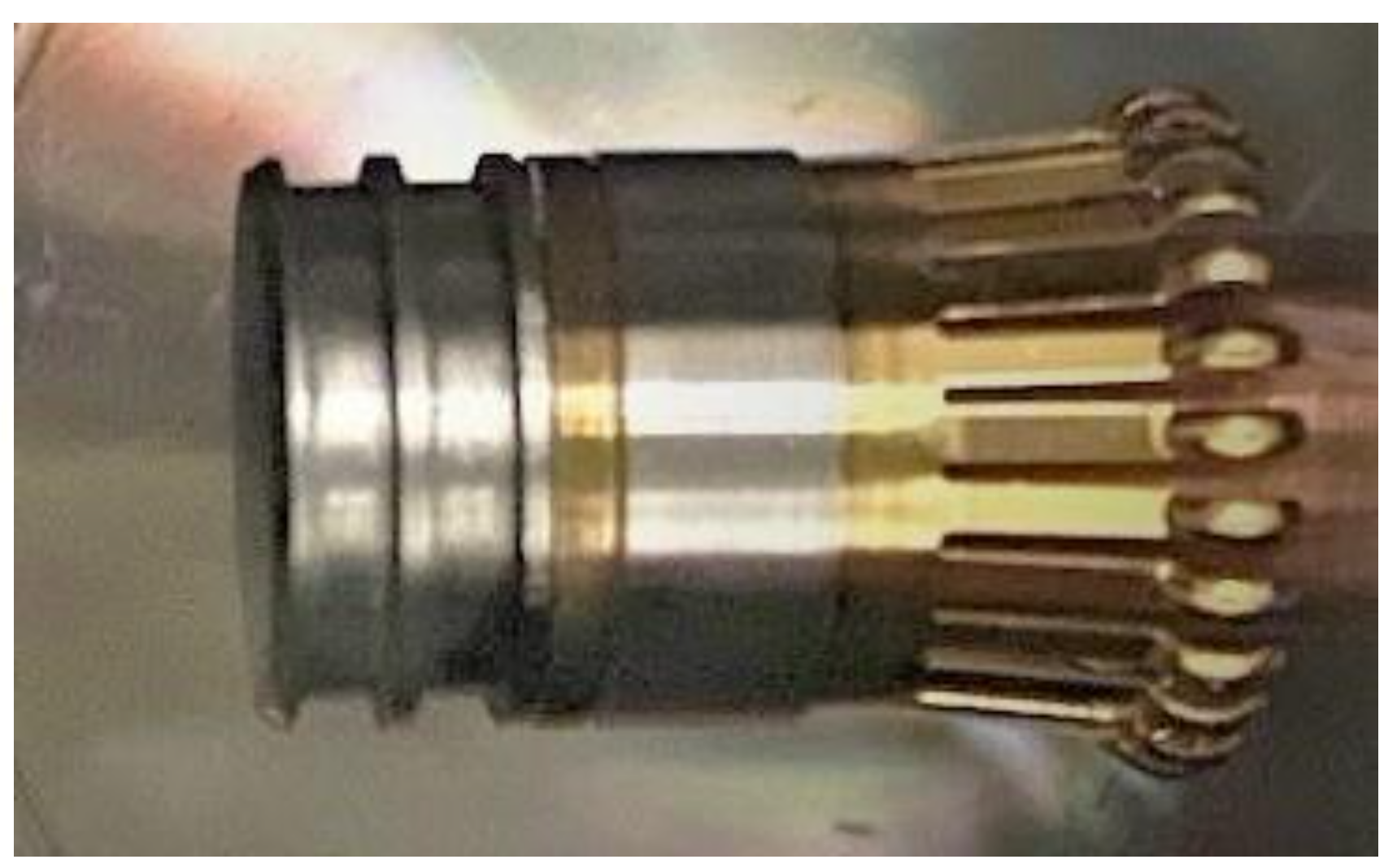}}
    \caption{(a) Cathode installed at the end of the cathode stalk. An asymmetric circular pattern can be seen here. (b) After the cathode was extracted from the gun, this image shows a gap that had opened up between the puck and the end effector.}
    \label{fig:cathode_assym}
\end{figure}

\begin{figure}
    \centering
    \subfloat{
    \includegraphics[height=3cm,width=.49\linewidth]{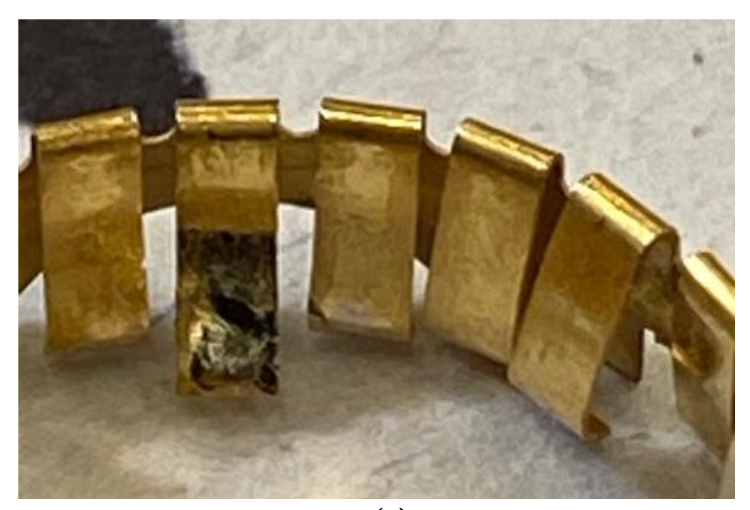}}
    \hfill
    \subfloat{%
    \includegraphics[height=3cm,width=.49\linewidth]{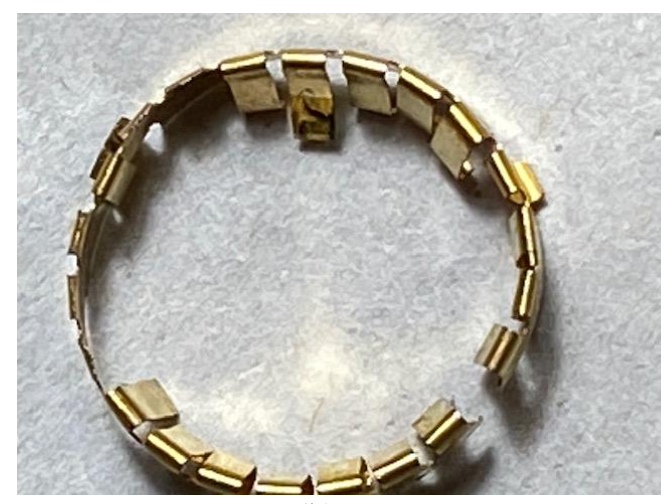}}
    \caption{(a) Spring finger element. One of the fingers clearly shows a burn mark. (b) Shows the plastic deformation of the spring finger being ``pulled forward''.}
    \label{fig:RF_burn}
\end{figure}

At the beginning  of the CeC experimental run 2021, the gun showed signs of significant performance degradation: rapid decay of the cathode quantum efficiency, appearance of new low-level multipacting barriers, increased radiation and LiHe consumption compared to the routine operation levels. These observations were then followed by the series of dramatic changes in the beam dynamics within the cavity. The measured axis of the beam propagation has shown a significant deviation from the straight path: the beam axis was tilted by 22~mrad horizontally indicating the corresponding tilt of the cathode surface (see \cite{petrushina2019measurements} for the detailed description of the gun axis measurement procedure). This fact was then confirmed by the direct observation of the asymmetry in the cathode placement within the cavity nose (see Fig.~\ref{fig:cathode_assym}). The reported behavior indicated a potential damage to the cathode delivery system and inevitable presence of the cavity surface contamination. Upon a detailed inspection, the damage to the cathode end effector was found to be the issue. When the puck was removed a close-up inspection found evidence that the manipulator had been damaged due to either poor or inattentive handling of the end effector or horizontal manipulator. As can be seen from Fig.~\ref{fig:RF_burn}, at least one of the RF spring fingers clearly showed a burn mark, which is the evidence that particulates were generated that may have escaped through the gap between the puck and the end effector registration surface.

\begin{figure}
    \centering
    \includegraphics[width=1\linewidth]{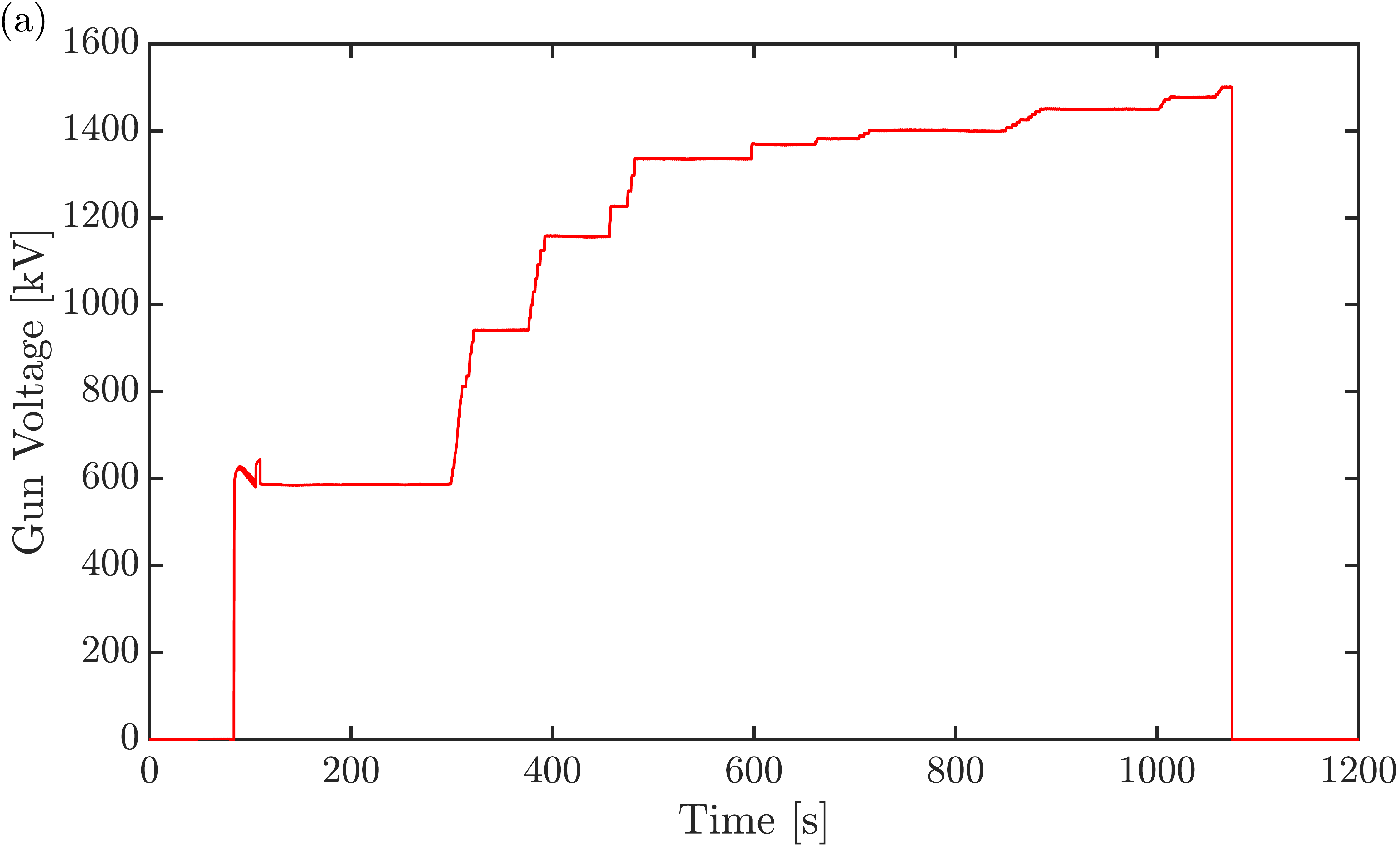}
    \includegraphics[width=1\linewidth]{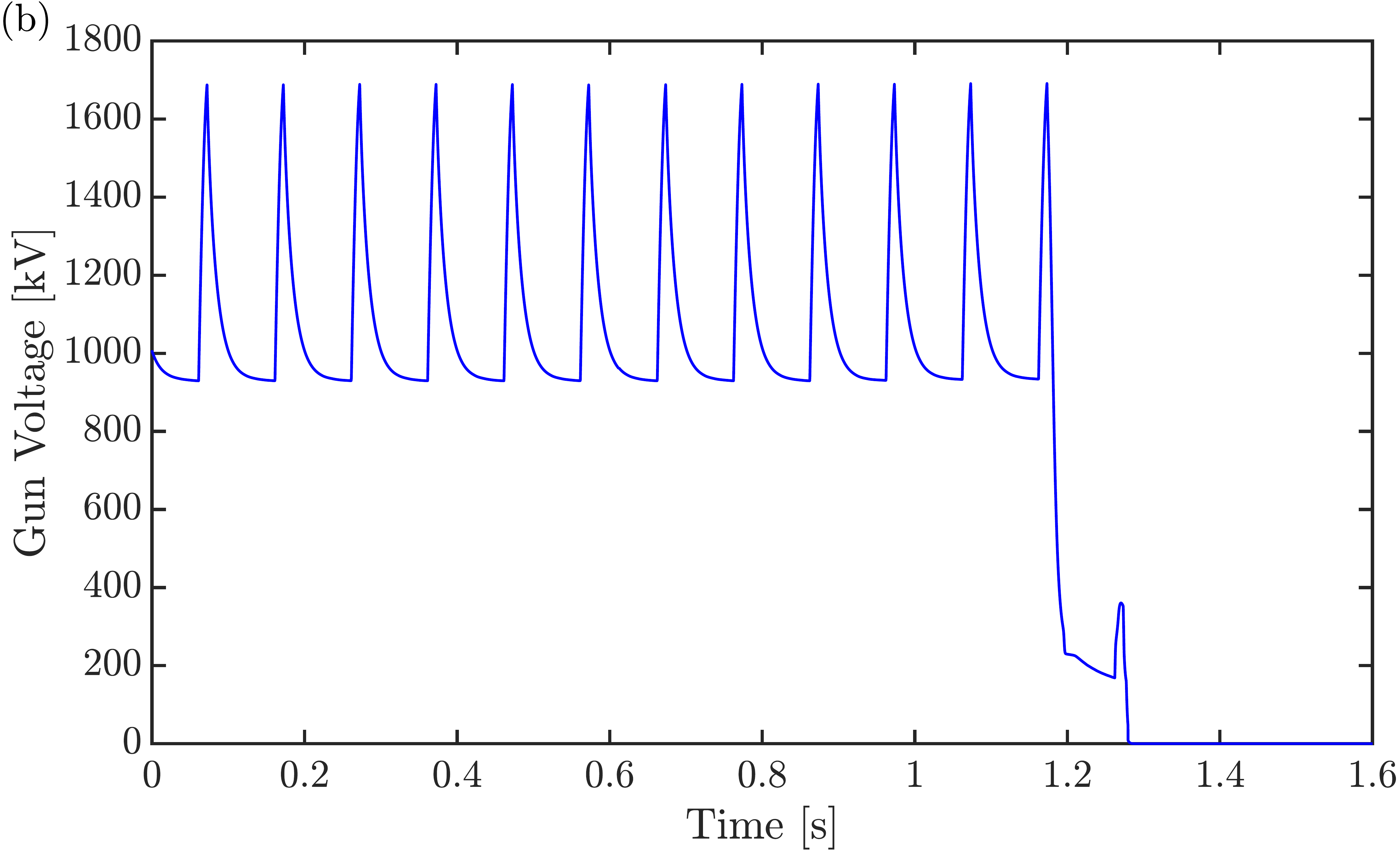}
    \caption{The highest achieved CW (a) and pulsed (b) cavity voltage after the first case of He conditioning.}
    \label{fig:HeCcase1}
\end{figure}

After the unsuccessful attempts to bring the cavity to the regular operational state by using CC and HPP, it was decided to proceed with He conditioning. The He conditioning of the CeC gun cavity typically adheres to the following procedure. First, before the He is introduced into the cavity, with the He leak valve and all the surrounding valves of the CeC system closed and the cathode puck retracted, the gun voltage must be brought up well above multipacting levels, usually at a baseline of 700-750~kV. The protection mechanism for monitoring the cavity voltage is then enabled to prevent the cavity from falling into multipacting for a prolonged period of time. The implemented protocol is registering the cavity voltage, and once the level is detected to be below the multipacting threshold (usually below 400~kV), the cavity is immediately turned off. The pulsing with 10~Hz repetition and pulse width of 1-2 ms is then introduced on top of the baseline similar to one shown in Fig.~\ref{fig:HeCcase1}b. Once the pulsing event is set, the He leak valve is opened to introduce He into the cavity at a desired pressure level, typically in the range of 2-5$\times10^{-6}$~Torr. The conditioning procedure consists of a gradual increase of the peak cavity voltage that can be achieved through either increase of the pulse length of the pulsing event or increase of the baseline level. Since the increase of the baseline level requires higher LiHe consumption, the conditioning is usually performed by the adjustment of the pulse length of the pulsing event. One needs to gradually increase the pulse width in the steps of $\sim$0.5~ms, stay at each level for an extended period of time (10-15 mins) and simultaneously monitor the gun voltage behavior, radiation levels, mass flow of liquid He, pressure in the gun cavity, and gauges of the ion pumps providing the pressure levels of the He used for conditioning.  During the HeC, He and the loose debris are being constantly pumped out of the cavity by a turbo pump. The overall process of HeC is rather delicate and sudden increases of the applied power could result in interruption by the cavity machine protection system (MPS).  If the gun trips, the protection mechanism turns the cavity off, one must close the He leak valve and wait for the vacuum to recover to $\sim1\times10^{-7}$ Torr before turning the system back on.

\begin{figure*}
    \centering
    \includegraphics[width=1\linewidth]{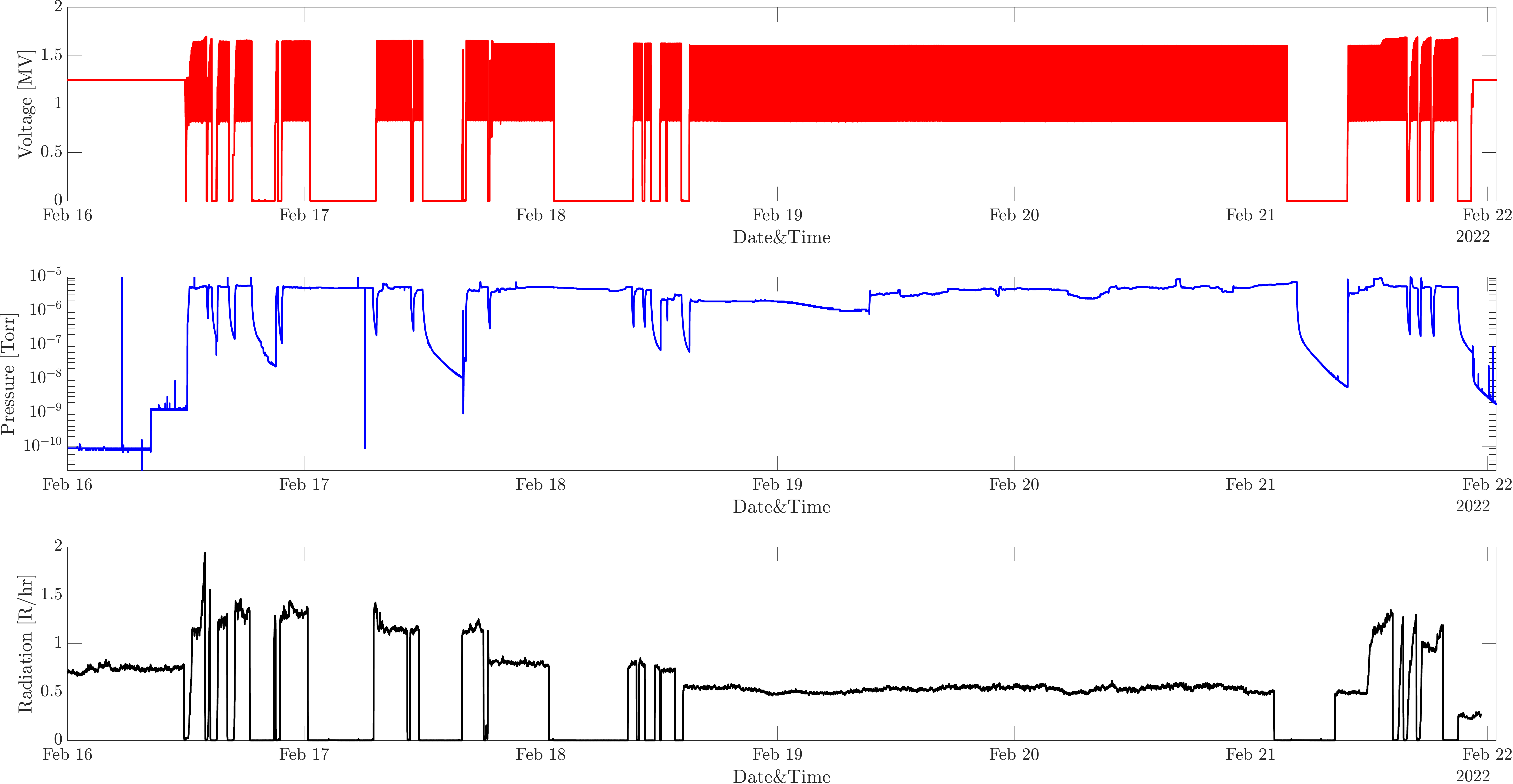}
    \caption{Gun voltage, pressure in the cavity and radiation levels during HeC.}
    \label{fig:HeCcase2}
\end{figure*}

After a couple of days of He conditioning, the cavity showed significant performance improvement: the gun reached the all-time highest voltage demonstrating 1.5~MV CW and 1.7~MV pulsed. The radiation level at the CeC operational level of 1.25~MV CW has also improved dramatically and reduced to $\sim$380 mR/hr with $\sim$2 g/sec LiHe consumption. Figure~\ref{fig:HeCcase1} shows the final test of the cavity operation in CW and pulsed modes after the completion of the He conditioning. The CW voltage limit is defined by the maximum available LiHe consumption of 8~g/sec. A quench-like limit occurs at 1.7 MV in pulsed mode as indicated by the steep drop of the cavity voltage in Fig.~\ref{fig:HeCcase1}b. The mechanism of capturing the cavity voltage when it falls below the MP level during the HeC pulsing procedure was implemented after this event, hence the appearance of the small spike in Fig.~\ref{fig:HeCcase1}b is an indication that the cavity wasn't turned off soon enough after the cavity tripped.

\subsection{He Conditioning: Case 2}

In February 2022, during a mundane cathode exchange, it was found that the cathode puck wasn’t fully engaged with the end effector of the cathode manipulator once the old cathode was extracted. While attempting to insert a replacement cathode, an obstacle was encountered inside the cathode stalk that prevented the team from inserting the cathode, unless a significant adjustment of the horizontal angle of the insertion arm was applied. The cathode was installed, but the cavity demonstrated an immediate performance degradation: although the MP barriers were surpassed without an issue, the cavity would not be able to exceed 600 kV, showing significant amount of dark current in the cathode area and several spots of visibly glowing emitters. An attempt to clean the cavity by using traditional CW or high-power processing (HPP) was successful at first, but eventually resulted in a cavity trip followed by significantly aggravated MP levels, that were impossible to overcome. The cathode was extracted, and after several attempts of HPP, it was decided to switch to HeC. Within the course of a few days, we performed HeC, and studied in detail how the amount of the delivered He affects the cleaning process. The evolution of the cavity voltage and respective cavity pressure and radiation levels are shown in Fig.~\ref{fig:HeCcase2}. The same procedure as in the previous case was applied, but the He pressure levels were adjusted in the range between 2$\times 10^{-6}$ to 6$\times 10^{-6}$ Torr based on the cavity behavior. Overall, HeC procedure was highly successful and resulted not only in the complete cavity recovery, but also in an all-time low X-ray radiation from the cavity ($\sim$250 mR/hr, compared to $\sim$380 mR/hr before the incident and $\sim$750 mR/hr right after the incident) at the operational voltage of 1.25 MV.

\section{Conclusions}

The development of cavity restoration techniques in application to the SRF photoinjectors is highly valuable for the future development and maintenance of the SRF electron sources. On the example of the BNL SRF photogun we have proven that He conditioning technique is highly successful in restoration of the cavity performance and allows to effectively remove contamination, reduce LiHe consumption and radiation levels, resulting in the improved perfromance of the gun cavity. 

\begin{acknowledgments}
Work is supported by Brookhaven Science Associates, LLC under Contract No. DEAC0298CH10886 with the U.S. Department of Energy, and the Department of Energy office of Nuclear Physics grant DE-FOA-0002670.
\end{acknowledgments}

\bibliography{main}

\end{document}